\documentclass[amsmath,amssymb,aps,reprint,pra]{revtex4-2}

\usepackage{algorithm}
\usepackage[noend]{algpseudocode}
\usepackage{mathtools}
\usepackage{placeins} % Provides \FloatBarrier
\usepackage{subfiles} % We use this package instead of \input, because we can then also compile subfiles.
\usepackage[hidelinks]{hyperref}
\usepackage{orcidlink}
\usepackage{datetime} % Needed for \newdate
\usepackage{makecell} % Needed for \makecell

\makeatletter
\def\algbackskip{\hskip-\ALG@thistlm}
\makeatother

\usepackage[acronym,hyperfirst=false]{glossaries}
\glsdisablehyper
\makeglossaries
\newacronym{cpu}{CPU}{central processing unit}
\newacronym{fft}{FFT}{fast Fourier transform}
\newacronym{fpga}{FPGA}{field-programmable gate array}
\newacronym{fso}{FSO}{free-space optical}
\newacronym{qber}{QBER}{quantum bit error rate}
\newacronym{qkd}{QKD}{quantum key distribution}
\newacronym{ram}{RAM}{random-access memory}
\newacronym{skr}{SKR}{secure-key rate}
\newacronym{sdn}{SDN}{software-defined network}
\newacronym{spad}{SPAD}{single-photon avalanche diode}

\begin{document}

\title{Dynamic rerouting and interruption resilience of quantum communication via single-photon-based resynchronization}

\author{\orcidlinki{Jan~Krause}{0000-0002-3428-7025}}
\email[]{jan.krause@hhi.fraunhofer.de}
\author{\orcidlinki{Stephanie~Renneke}{0009-0001-5910-5048}}
\author{Jonas~Hilt}
\author{\orcidlinki{Oliver~Peters}{0009-0005-1081-4723}}
\author{\orcidlinki{Peter~Hanne}{0009-0005-6914-0233}}
\author{\orcidlinki{Andy~Schreier}{0000-0002-8424-3899}}
\author{\orcidlinki{Ronald~Freund}{0000-0001-9427-3437}}
\author{\orcidlinki{Nino~Walenta}{0000-0001-7243-0454}}

\affiliation{Fraunhofer Institute for Telecommunications, Heinrich-Hertz-Institute, HHI, 10587 Berlin, Germany}

\newdate{date}{20}{08}{2025}
\date{\displaydate{date}}
%\date{\today}

\begin{abstract}
    We present a resynchronization method for quantum key distribution (QKD) systems that enables rapid and reliable recovery from interruptions of the quantum channel and changes of its optical path length.
    By periodically transmitting short fixed pulse patterns over the quantum channel, our approach achieves swift clock offset recovery, typically within a few hundred milliseconds.
    We implemented this method in our time-bin-phase BB84 QKD system, demonstrating successful resynchronization after multi-minute channel interruptions and fiber length changes exceeding $100\,\mathrm{km}$.
    The method can be retrofitted to existing systems via a software upgrade and without hardware changes, allowing for broad applicability.
    In total, the resynchronization method significantly enhances QKD system resilience and allows for reliable operation in challenging environments such as dynamically routed optical networks, i.e., software-defined networks, and free-space optical links with mobile nodes.
\end{abstract}

\maketitle
\section{\label{sec:introduction}Introduction}

\begin{figure}[b]
    \centering
    \includegraphics[width=\linewidth]{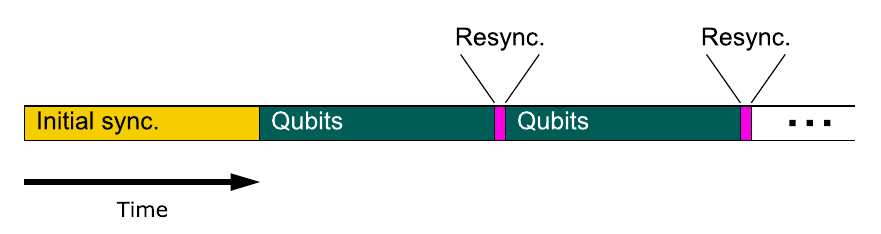}
    \caption{\label{fig:sync_order}
    Transmission sequence.
    After the initial synchronization, qubit blocks and short resynchronization blocks are alternately transmitted over the quantum channel.
    As an example, a typical configuration uses $\sim 1\,\mathrm{s}$ qubit blocks and $\sim 10\,\mathrm{ms}$ resynchronization blocks.
    }
\end{figure}

\Gls{qkd} provides symmetric cryptographic keys to two distant parties
\cite{
    bennettQuantumCryptographyPublic1984,
    gisinQuantumCryptography2002,
    pirandolaAdvancesQuantumCryptography2020}.
It allows for provable protocol security, a unique feature that remains to be established for conventional, purely algorithmic alternatives, which rely on unproven assumptions about the computational complexity of certain mathematical problems.

However, \gls{qkd} requires the exchange of single-photon-encoded symbols via a direct optical connection, called quantum channel, between the distant parties.
Channel interruptions lead to a drift of the free-running clocks of Alice and Bob, and fiber length changes add a length-difference-dependent additional offset.
Both effects cause a loss of synchronization between sender and receiver, typically requiring a system restart.

In practical \gls{qkd} operating regimes, the detection stream at the receiver is extremely sparse: Bob typically detects one photon for every $10$ to $10^7$ symbols transmitted by Alice
\cite{boaronSecureQuantumKey2018}.
At the same time, the quantum bit error rate (QBER) can climb to around 11\,\%
\cite{gottesmanSecurityQuantumKey2004},
far above the $\sim 10^{-9}$ (or lower) bit-error levels routinely achieved in classical fiber‑optic systems
\cite{agrawalFiberopticCommunicationSystems2010}.
This combination of severe loss and comparatively high error violates the assumptions underpinning conventional telecom synchronization, necessitating dedicated synchronization strategies tailored to the quantum setting.

In this work, we introduce a single-photon-based resynchronization method, for swift clock offset recovery following channel interruptions or changes of the optical path length.
Our method works by periodically transmitting short blocks of fixed pulse patterns that are interleaved into the qubit stream, see Fig.~\ref{fig:sync_order}.
The method is easy to implement and can be retrofitted to prepare-and-measure QKD systems via a software upgrade and without hardware changes.

We implemented the resynchronization method in our time-bin-phase BB84 QKD system
\cite{krauseFlexibleRealTimeQuantum2024},
demonstrating reliable real-time operation.
Clock offsets were recovered within a few hundred milliseconds, even after prolonged quantum channel interruptions of multiple minutes and fiber length changes of more than 100 km.

\begin{figure*}
    \centering
    \includegraphics[width=\linewidth]{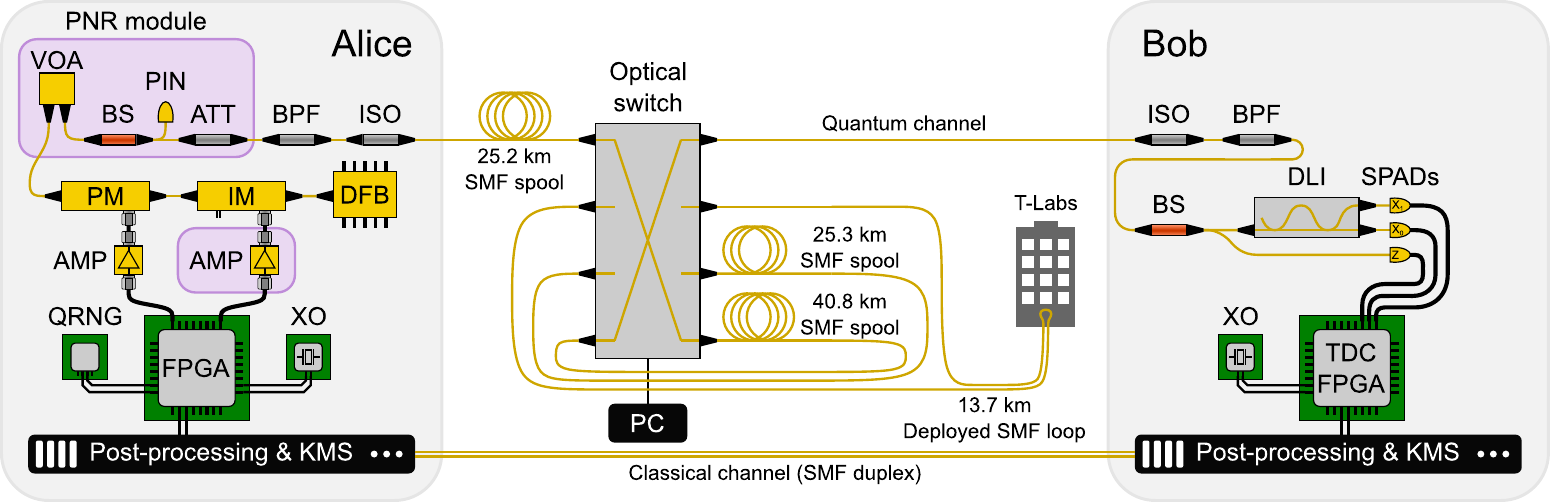}
    \caption{\label{fig:setup}
    System design and field experimental setup.
    The system operates at a qubit transmission rate of $625\,\mathrm{MHz}$, using two $800\,\mathrm{ps}$ time-bins per qubit.
    The quantum channel is used for the transmission of qubits, syntonization (clock frequency recovery), initial synchronization \cite{krauseClockoffsetRecoverySublinear2025} (clock offset recovery), and for resynchronization.
    The Ethernet connection is used for system control and post-processing.
    Both connections may be interrupted and change their optical path length during live operation.
    Thin (thick) black lines depict digital (analog) electrical signals.
    AMP: Electrical amplifier;
    ATT: Fixed attenuator;
    BPF: Band-pass filter;
    BS: Beam splitter;
    DFB: Distributed-feedback laser;
    DLI: Delay-line interferometer;
    FPGA: Field-programmable gate array;
    IM: Intensity modulator;
    ISO: Optical isolator;
    KMS: Key management system;
    PC: Personal computer;
    PIN: Photodiode;
    PNR: Photon-number regulation;
    PM: Phase modulator;
    QRNG: Quantum random number generator;
    SMF: Single-mode optical fiber;
    SPAD: Single-photon avalanche diode;
    TDC: Custom time-to-digital converter;
    T-Labs: Telekom Innovation Laboratories;
    VOA: Variable optical attenuator;
    XO: Free-running low-cost quartz crystal oscillator.
    }
\end{figure*}

Hence, our method greatly increases usage flexibility and enables system operation in the most challenging environments, such as dynamically routed optical networks and free-space optical links.
When integrating \gls{qkd} into software-defined networks (SDNs)
\cite{
    martinMadQCIHeterogeneousScalable2024,
    aliaDynamicDVQKDNetworking2022,
    simSoftwareDefinedNetworkingOrchestration2023},
our method enables changing optical quantum channel routes in real-time.
This capability enhances system flexibility, e.g., to handle varying traffic and routing demands
\cite{
    bhatiaDynamicSecurityAwareResource2025,
    selentis-boulntadakisRelayedVsSwitched2025}.

This paper is organized as follows.
First, the QKD system, in which the resynchronization was implemented, is introduced in Sec.~\ref{sec:setup}.
Resynchronization is then described in detail in Sec.~\ref{sec:sync_method}.
Experimental results are presented in Sec.~\ref{sec:experiment}, and a conclusion is given in Sec.~\ref{sec:conclusion}.

A derivation of the success probabilities for the resynchronization method is provided in Appendix~\ref{app:parameter_choice}, leading to criteria for the suitable choice of the method parameters.
First considerations of the impact of our method on QKD security with regard to potential side channels are provided in Appendix~\ref{app:security}.
The nomenclature is summarized in Appendix~\ref{app:nomenclature}.

\section{\label{sec:setup}Experimental setup}

The presented resynchronization method was implemented in our QKD system with real-time post-processing, see Fig.~\ref{fig:setup}, which uses the \mbox{1-decoy} time-bin-phase BB84 protocol
\cite{ruscaFinitekeyAnalysis1decoy2018}.
The system transmits qubits at a rate of $625\,\mathrm{MHz}$, using two $800\,\mathrm{ps}$ time-bins per qubit.

The transmitter (Alice) prepares qubits by tailoring the output of a $1546.92\,\mathrm{nm}$ continuous-wave distributed-feedback (DFB) laser, using an intensity modulator (IM) and a phase modulator (PM).
This results in optical pulses with full-width at half maximum (FWHM) of approx. $200\,\mathrm{ps}$.
Entropy for the random qubit generation is extracted from a quantum random number generator (QRNG, ID Quantique IDQ20MC1-T) and expanded using an AES-CTR scheme, allowing for precise adjustment of the probabilities for the choice of bases and decoy levels.
Qubit detection is performed by the receiver (Bob) using free-running single-photon avalanche diodes (SPADs, ID Quantique IDQube), operated at 20\,\% quantum efficiency and $20\,\mathrm{\mu s}$ dead-time.
The system uses a custom-built time-to-digital conversion (TDC) module, which is based on a field-programmable gate array (FPGA) and achieves a timing resolution of $100\,\mathrm{ps}$.

Only two connections are needed between Alice and Bob.
The optical quantum channel is used for the transmission of qubits, syntonization (clock frequency recovery), initial synchronization
\cite{krauseClockoffsetRecoverySublinear2025} (clock offset recovery), and for resynchronization.
The classical channel is used for system control, monitoring, and post-processing.
The requirements for the classical channel are very low, as it does not need to be transmitted optically, may suffer from unknown latencies of up to hundreds of milliseconds, and can be routed via a different path than the quantum channel.
It can be seen as the task of the resynchronization to allow for a similar flexibility for the quantum channel.

The system uses a qubit-based method for \textit{syntonization}, i.e., clock-frequency recovery, which is achieved entirely in software during post-processing.
This method allows sender and receiver to use independent free-running low-cost quartz crystal oscillators, operating without temperature stabilization.

For the syntonization, the single-photon detection timestamps are processed in blocks with a few hundred detections each.
For each block, the clock drift is estimated and compensated by a linear approximation, such that a contrast criterion is maximized.

We applied interruptions and fiber length changes to the quantum channel, which are representative of those encountered in real deployed networks.
To achieve this reproducibly, we used a 4x4 optical switch (Gezhi S+C+L band FWDM module), that was digitally-controlled by a separate PC.
This PC had no connection or synchronization with the QKD system, such that the resynchronization had to rely entirely on the single-photon detections, representative of a real-world scenario.
The setup allowed to configure quantum channel fiber lengths of 25.2, 39.0, 50.5, and 66.0~km, with total losses of 7.4, 16.8, 14.4, and 19.1~dB, respectively.
Each pass through the optical switch accounted for approx. 1.75~dB of loss.

\section{\label{sec:sync_method}Synchronization}

\begin{figure}
    \begin{algorithm}[H] % The [H] is needed for this to work with RevTex
        \caption[justification=justified]{\label{alg:resync_eval}
        Offset recovery algorithm.
        The algorithm is designed such that the evaluation finishes very quickly for the most likely case, i.e., an offset $\Delta = 0$.
        $\cdot \gg 1$: Bit shift by 1, equal to integer division by 2.
        }
        \hspace*{\algorithmicindent} \textbf{Input:} $\mathcal{P} = (\mathcal{P}_1, \dots, \mathcal{P}_{N_\mathrm{r}})$ \Comment{Trans. resync. pattern} \\
        \hspace*{\algorithmicindent} \textbf{Input:} $\mathcal{D}^{(s)} = (\mathcal{D}^{(s)}_1, \dots, \mathcal{D}^{(s)}_{N_\mathrm{d,i}})$ \Comment{Detections} \\
        \hspace*{\algorithmicindent} \textbf{Input:} $\Delta_\mathrm{max}$ \Comment{Max. offset to test in time-bin prec.} \\
        \hspace*{\algorithmicindent} \textbf{Input:} $t$ \Comment{Cross-correl. acceptance threshold} \\
        \hspace*{\algorithmicindent} \textbf{Input:} $\tau$ \Comment{Timebin duration} \\
        \hspace*{\algorithmicindent} \textbf{Output:} $\delta_\mathrm{tot}^{(\mathrm{s})}$ \Comment{Found offset}
        \begin{algorithmic}[1]
        \Function{FindOffset}{$\mathcal{P}, \mathcal{D}, \Delta_\mathrm{max}, t, \tau$}
            \State $\delta_\mathrm{align}^{(\mathrm{s})} \gets \Call{FindAlignmentOffset}{\mathcal{D}^{(s)}, \tau}$
            \State $\mathcal{D}^{(s)} \gets \Call{AlignDetections}{\mathcal{D}^{(s)}, \delta_\mathrm{align}^{(\mathrm{s})}}$
            \State $\mathcal{D} \gets \Call{ConvertToTimebinPrecision}{\mathcal{D}^{(s)}, \tau}$
            \State $\mathcal{D} \gets \Call{RemoveMargins}{\mathcal{D}, N_\mathrm{r}, \Delta_\mathrm{max}}$
            \State $T \gets \lceil t \times \Call{Size}{\mathcal{D}} \rceil$ \Comment{Abs. threshold}
            \For{$n \gets 0$ to $2 \Delta_\mathrm{max}
            $} \Comment{Iterate over all offsets}
                \State $\Delta \gets \Call{MapNatToIntBijective}{n}$
                \State $C \gets \Call{CalcCrossCorrelation}{\mathcal{P}, \mathcal{D}, \Delta}$
                \If{$C > T$} \Comment{Check acceptance criterion}
                    \State $\delta_\mathrm{tot}^{(\mathrm{s})} \gets \Delta \times \tau + \delta_\mathrm{align}^{(\mathrm{s})}$
                    \State \Return $\delta_\mathrm{tot}^{(\mathrm{s})}$
                \EndIf
            \EndFor
            \State \Return 0 \Comment{In case no offset qualifies in line 10}
        \EndFunction
        \Statex
        \Function{MapNatToIntBijective}{$n$}
            \If{$(n \mod 2) = 0$}
                \State \Return $- (n \gg 1)$
            \Else
                \State \Return $(n \gg 1) + 1$
            \EndIf
        \EndFunction
        \Statex
        \Function{CalcCrossCorrelation}{$\mathcal{P}, \mathcal{D}, \Delta$}
            \State $N \gets 0$ \Comment{Counter for coinciding pulses}
            \For{$i \gets 0$ to $\Call{Size}{\mathcal{D}}$}
                \State $k \gets i - \Delta$ \Comment{Apply offset}
                \State $N \gets N + \mathcal{P}_k$ \Comment{Incr. counter, if pulses coincide}
            \EndFor
            \State $C \gets 2 N - \Call{Size}{\mathcal{D}}$
            \State \Return $C$
        \EndFunction
        \end{algorithmic}
    \end{algorithm}
\end{figure}

Requirements for the initial synchronization and resynchronization are quite different.
Due to the relaxed requirements to the classical channel with respect to latencies, the initial synchronization needs to identify temporal offsets of up to $\sim 100\,\mathrm{ms}$.
Hence, the correct offset needs to be found out of $\sim 10^8$ possibilities.
In contrast, for resynchronization, it is most likely that the offset does not change in between consecutive resynchronization blocks.
Only in a small fraction of cases, e.g., after a channel interruption or fiber length change, the system needs to identify a non-zero additional offset.
In most scenarios, fiber length changes are expected to remain within $\pm 100\,\mathrm{km}$, corresponding to temporal offsets $\lesssim 1\,\mathrm{ms}$ ($\sim 10^6\,\mathrm{timebins}$).
This is summarized in Table~\ref{tab:sync_requirements}.

\subsection{Initial synchronization}

For initial \textit{synchronization}, i.e., the initial clock offset recovery, the system uses iQSync
\cite{krauseClockoffsetRecoverySublinear2025}.
Therefor, Alice sends a coarse marker to Bob via the classical channel, and transmits a dedicated pattern of \textsf{Z}-basis qubits via the quantum channel.
Since iQSync allows for the resource-efficient recovery of clock offsets of up to hundreds of milliseconds, the timing requirements for the classical channel are relaxed.
To recover the clock offset, Bob evaluates the single-photon detection timestamps according to Ref.~\cite{krauseClockoffsetRecoverySublinear2025}.

\begin{table}[t]
    \caption{\label{tab:sync_requirements}
    Comparison of synchronization requirements.
    The initial offset is potentially very large and a priori knowledge about the most probable offset is hard to use.
    For the resynchronization, an additional offset of zero timebins is most likely.
    This a priori knowledge can be used.
    }
    \begin{ruledtabular}
        \begin{tabular}{l|cc}
             & \makecell{Initial \\ synchronization} & \makecell{Resynchro- \\ nization} \\
            \colrule
            Max. offset         & $\approx 100 \, \mathrm{ms}$                         & $\approx 100\,\mathrm{km}$ \\%[1.5ex]
            Equivalent search space         & $\approx 10^8\,\mathrm{timebins}$                    & $\approx 10^6\,\mathrm{timebins}$ \\
            Avg. eval. runtime   & $\lesssim 10 \, \mathrm{s}$       & $\lesssim 1 \, \mathrm{ms}$ \\
        \end{tabular}
    \end{ruledtabular}
\end{table}

\begin{figure*}[t]
    \includegraphics[width=\linewidth]{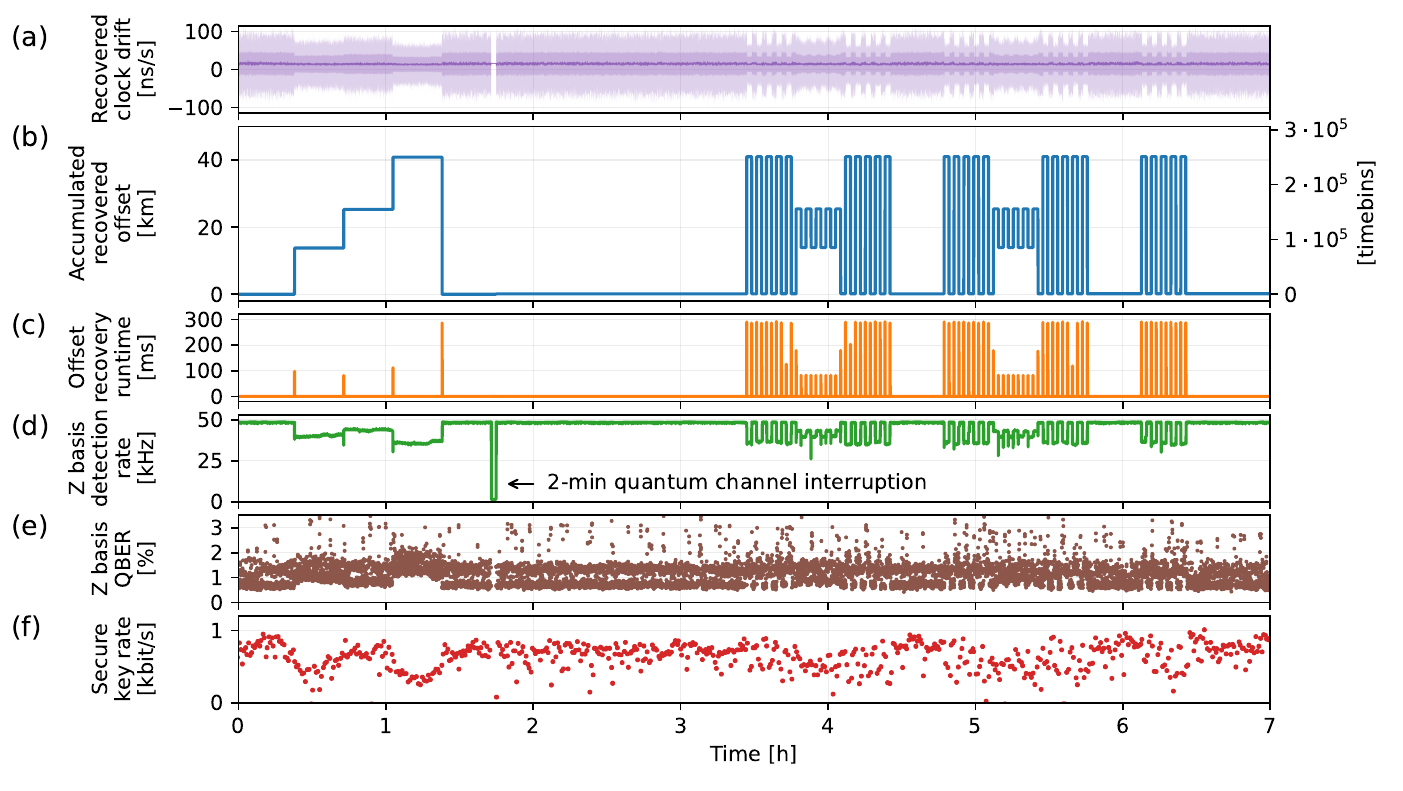}
    \caption{\label{fig:results}
    Experimental results.
    (a) depicts the applied qubit-based clock drift correction. The light, medium, and dark lines show the extreme values, standard deviation, and average value for a two-second sliding window, respectively.
    Lower detection rates lead to an increased averaging effect of the obtained drift factors of the fixed-length syntonization block.
    This effect leads to the false impression of higher clock stability for higher channel attenuations.
    (b) shows the accumulated offset as it was recovered by Bob based on the resynchronization blocks, demonstrating its reliability.
    The 2-min quantum channel interruption introduced an additional offset of only $800\,\mathrm{ps}$ (not discernible in the figure).
    (c) shows the recovery runtime of the \texttt{C++} implementation, which was consistently approx. $50\,\mathrm{\mu s}$ for blocks without offset, i.e., $\Delta = 0$.
    (d), (e), and (f) depict the $\mathsf{Z}$-basis detection rate, $\mathsf{Z}$ basis QBER, and secure-key rate, respectively.
    The system was operated without issues for 7 hours.
    At the beginning, the 4 fiber lengths were configured for $20\,\mathrm{min}$ each, followed by a 2-min channel interruption.
    Afterwards, 70 fiber length changes of up to 40.8~km were applied.
    The offset was successfully recovered in all instances.
    The qubit block length was set to $2^{28}$ timebins ($214.7\,\mathrm{ms}$), and the resynchronization block length to $2^{25}$ timebins ($26.8\,\mathrm{ms}$).
    A maximum of $N_\mathrm{d} = 500$ detections was used for the evaluation of each resynchronization block.
    The acceptance threshold was set to $t = 0.5$.
    }
\end{figure*}

\subsection{Resynchronization}

For \textit{resynchronization}, the sender alternately transmits qubit blocks and resynchronization blocks, see Fig.~\ref{fig:sync_order}.
These resynchronization blocks consist of a fixed pseudo-random pattern of \textsf{Z}-basis qubits.
Both the size of qubit blocks and resynchronization blocks can be configured independently.

After reception of each resynchronization block, Bob recovers a potential offset by calculating the cross-correlation between received single-photon detection timestamps and the known pattern transmitted by Alice.
Offsets are tested consecutively, starting with the most likely offset of zero timebins, i.e., $\Delta = 0$.
This approach minimizes the computational load for $\Delta = 0$, leading to a swift evaluation during undisturbed system operation.

The recovery algorithm is shown in pseudo-code in Algorithm~\ref{alg:resync_eval}.
First, the detections are aligned to the time-bin center and converted to time-bin precision (lines 2--4).
To avoid out-of-range indexing in line 23, detections within the first and last $\Delta_\mathrm{max}$ timebins of the block are removed (line 5).
Here, $\Delta_\mathrm{max}$ refers to the configured maximum tested offset in units of time-bins.
Offsets are then tested in a simple loop (line 7) by calculating the cross-correlation between the transmitted pattern and the detection timestamps (line 9).
If the cross-correlation exceeds the configurable threshold $t$ (line 10), the tested offset $\Delta$ is accepted, terminating the evaluation.
Offsets $\Delta$ are tested in the order $0, \pm 1, \pm 2, \dots, \pm \Delta_\mathrm{max}$ (line 8).
To optimize the evaluation runtime and reduce branch misprediction penalties, the cross-correlation is calculated without if-clauses (lines 19--25).

The resynchronization method provides free choice of four parameters: the qubit block size $N_\mathrm{q}$, the resynchronization block size $N_\mathrm{r}$, the acceptance threshold $t$, and the maximum tested offset $\Delta_\mathrm{max}$.
This freedom allows for a flexible adaptation to the channel conditions to achieve the required success probability, while keeping a small \gls{skr} penalty due to the lower qubit transmission duty cycle.
The success probabilities of our method are derived in Appendix~\ref{app:parameter_choice}, which leads to criteria for the choice of the four parameters.

\section{\label{sec:experiment}Experiment}

We experimentally validated our method using the QKD system described in Sec.~\ref{sec:setup}, demonstrating reliable and swift clock recovery within less than 1~s after channel interruptions and fiber length changes exceeding 40~km.

On Alice's side the signal generation for initial synchronization and resynchronization were implemented in the FPGA.
For Bob, the recovery algorithms for syntonization, initial synchronization, and resynchronization were implemented in a \texttt{C++} program, also executed in real-time using a single thread of a recent server-grade central processing unit (CPU; Intel\textsuperscript{\textregistered} Xeon\textsuperscript{\textregistered} Silver 4314 CPU, 2.40\,GHz), featuring L1d, L2, and L3 cache sizes of 48, 1,280, and 24,576\,kB, respectively.
% 3 * 2**14 byte 48 kB
% 5 * 2**18 byte 1280 kB
% 3 * 2**23 byte 24'576 kB

The qubit block size was set to $N_\mathrm{q} = 2^{28}$ timebins ($214.7\,\mathrm{ms}$), the resynchronization block size to $N_\mathrm{r} = 2^{25}$ timebins ($26.8\,\mathrm{ms}$), the cross-correlation acceptance threshold to $t=0.5$, and the maximum tested offset to $\Delta_\mathrm{max} = 10^6$ timebins, corresponding to a distance of $163 \, \mathrm{km}$ in standard single-mode fiber, see the cross in Fig.~\ref{fig:contour_plots}(b).
These values were chosen based on the analysis provided in Appendix~\ref{app:parameter_choice}.

To reduce memory access latencies to a minimum, the \texttt{C++} implementation at Bob used a dense format to store the resynchronization pattern, using one bit per timebin.
With a size of 4,096\,kB, the pattern $\mathcal{P}$ was small enough to completely fit into the CPU's L3 cache.
The detections vector $\mathcal{D}$ with a size of 500 detections, requiring 2\,kB when encoded as 32-bit integers, was small enough to even fit into the L1d cache.
Together, this potentially eliminated the need for random-access memory (RAM) usage during the evaluation, for which access latencies are typically approx. 10 times (100 times) longer compared to L3 (L1d) cache access \cite{greggSystemsPerformanceEnterprise2021}.

As shown in Fig.~\ref{fig:results}, the system ran uninterrupted for 7\,h.
During a test sequence at the beginning, the quantum channel was set to all four fiber lengths for 20\,min each.
The offset resulting from the drift of the free-running quartz oscillators during a 2-min channel interruption was successfully recovered within 1\,ms.

Another 70 fiber length changes of up to 40.8\,km were applied afterwards, using fiber spools and deployed fiber.
All resulting offsets were recovered without error and in less than 1\,s.
The resynchronization recovery runtime per fiber-length change was consistently measured at approx. 7\,ms/km.
In all cases with no additional offsets, the evaluation took approx. $50\,\mathrm{\mu s}$.

Furthermore, we tested fiber-length changes of up to 120\,km, which were also successfully recovered.

\section{\label{sec:conclusion}Conclusion}

We have presented a method that allows for the resynchronization of a QKD system after long quantum channel interruptions and substantial fiber length changes, without the need for a system restart.
Implemented in our time-bin-phase BB84 QKD system, we demonstrated swift and reliable clock offset recovery following quantum channel interruptions of up to 2\,min and fiber length changes of more than 100\,km.

The resynchronization method can be retrofitted to existing systems via a software upgrade and without hardware changes.
The method is well suited for implementation in low-level hardware, like FPGAs, since only simple hardware instructions, e.g., no floating-point operations, are used.
Furthermore, due to its conception, the method adds almost no additional system complexity.
Specifically, no additional subsystem interdependencies are introduced, as is, e.g., the case for other synchronization methods
\cite{cochranQubitBasedClockSynchronization2021}.

The impact on the SKR, due to the time required for the transmission of the resynchronization blocks, can be kept negligible.
For example, a $\mathsf{Z}$-basis detection rate of $50\,\mathrm{kHz}$ allows for resynchronization blocks as short as 10\,ms.
Combined with a resynchronization transmission interval of 10\,s, the SKR is only reduced by 0.1\,\% during undisturbed operation.

Our method can enhance system resilience in many scenarios.
In dynamically routed optical networks, it allows for the need-adapted change of optical routes, e.g., in case of fiber damage, or varying traffic demands.
For free-space optical links, it enables reliable QKD operation even in challenging environments, such as those involving mobile nodes, where channel interruptions can be caused by tracking issues, turbulence, or cloud coverage.
Furthermore, the resynchronization method can be used to recover from synchronization loss caused by other issues, such as errors in the syntonization step.
For qubit-based syntonization, such errors might occur due to statistical fluctuations of the single-photon detections, or due to phase instabilities in the free-running local clocks.

In summary, the resynchronization method is easy to implement and adds little system complexity, while significantly enhancing QKD system resilience and allowing for reliable operation in challenging environments.

\section{\label{sec:acknowledgements}Acknowledgments}

We thank Ralf-Peter Braun from Deutsche Telekom Laboratories (T-Labs) for providing access to the fiber loop.
Furthermore, we thank Abdelrahmane Moawad and Johannes Fischer for providing the optical switch and support with its operation.
An internal large language model was used during writing to polish and lightly edit some passages.
This research was conducted within the scope of the project QuNET, funded by the German Federal Ministry of Research, Technology and Space (BMFTR) in the context of the federal government's research framework in IT-security “Digital. Secure. Sovereign.”.

\section{\label{sec:author_contributions}Author Contributions}

J.K. conceived the resynchronization method and implemented the recovery algorithm for Bob, J.H. implemented it in Alice's FPGA.
J.K, S.R, and N.W. conceived the qubit-based clock recovery method, S.R. and J.K. implemented it.
O.P. built the custom clock boards.
J.H. and P.H. built the timetagger.
J.K. conducted the analytical success probability analysis, performed and evaluated the experiments, and wrote the manuscript.
A.S. improved the hardware design for the used second iteration of the system.
R.F. and N.W. acquired the funding.
N.W. guided and supervised the research.

\appendix

\begin{figure*}[ht]
    \includegraphics[width=\linewidth]{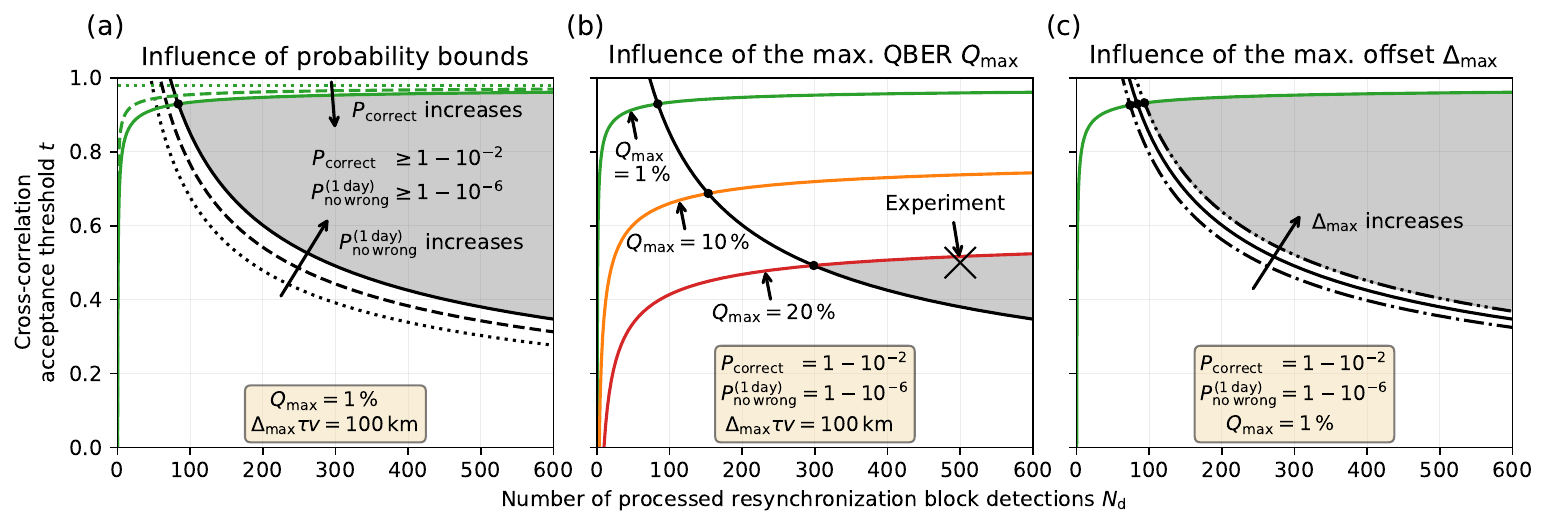}
    \caption{\label{fig:contour_plots}
    Success probability and parameter choice.
    One can freely choose the lower bound for the probability for not accepting a wrong offset over the course of one day, $P_\mathrm{no\,wrong}^{(\mathrm{1\,day})}$, the lower bound for the probability for accepting the correct offset with the first resynchronization block, $P_\mathrm{correct}$, the maximum tolerable QBER, $Q_\mathrm{max}$, and the maximum tested offset, $\Delta_\mathrm{max}$.
    The solution space for $N_\mathrm{d}$ and $t$ with regards to these constraints is shaded gray.
    The black dots depict the optimal solution, i.e., the one with the smallest possible $N_\mathrm{d}$.
    (a) shows how increasing $P_\mathrm{correct}$ and $P_\mathrm{no\,wrong}^{(\mathrm{1\,day})}$ shrinks the solution space.
    The black dotted (dashed, solid) contour stands for a probability of $0.5$ ($10^{-3}$, $10^{-6}$) for accepting one wrong offset per day, assuming the transmission of one resynchronization block per second, and a max. tested fiber length change of $\Delta_\mathrm{max} \tau v = 100\,\mathrm{km}$, where $\tau = 800 \, \mathrm{ps}$ is the time-bin duration, and $v \approx 2.04 \times 10^8 \, \mathrm{m/s}$ is the speed of light in the fiber.
    The green dotted (dashed, solid) contour stands for a probability of $0.5$ ($10^{-1}$, $10^{-2}$) for not accepting the correct new offset with one resynchronization block evaluation, given that the offset has changed.
    Note that in cases where the correct offset is not identified, it can still be recovered by the following resynchronization blocks.
    (b) shows how an increased QBER tolerance shrinks the solution space.
    Also see Eq.~\eqref{eq:threshold_condition}.
    The cross marks the parameters used for the experiment, see Sec.~\ref{sec:experiment}.
    (c) shows how increasing the maximum tested offset $\Delta_\mathrm{max}$ shrinks the solution space, with the dash-dotted (solid, dash-dot-dotted) line representing a maximum tested offset of $1\,\mathrm{km}$ ($10^2\,\mathrm{km}$, $10^4\,\mathrm{km}$).
    In total, a parameter choice near $N_\mathrm{d} = 300$, $t = 0.5$, and $\Delta_\mathrm{max} \tau v = 100\,\mathrm{km}$ covers a wide range of scenarios and should be suitable for most applications.
    }
\end{figure*}

\section{\label{app:parameter_choice}Success probabilities and parameter choice}

In this section, the resynchronization success probability is calculated in terms of the method parameters and the channel properties, allowing for the choice of suitable parameters in any scenario.

Our resynchronization method allows for the configuration of four parameters: The qubit block length in the number of time-bins, $N_\mathrm{q}$, the resynchronization block length in the number of time-bins, $N_\mathrm{r}$, the cross-correlation acceptance threshold, $t$, and the maximum tested offset in the number of time-bins, $\Delta_\mathrm{max}$.

We first derive the probability for not accepting a wrong offset, $P_\mathrm{no\,wrong}$, and the probability for accepting the correct offset, $P_\mathrm{correct}$.
Therefor, we need to describe the probability distribution of the cross-correlation for one resynchronization block.
The cross-correlation between the transmitted pattern $\mathcal{P}$ and the detection timestamps $\mathcal{D}$ for an applied offset of $\Delta$ is given by
\begin{align}
\mathrm{corr}_{\mathcal{P}, \mathcal{D}}(\Delta) = \frac{2 N - N_\mathrm{d}}{N_\mathrm{d}} \, ,
\end{align}
where $N$ is the number of detections coinciding with transmitted pulses for a tested offset $\Delta$, and $N_\mathrm{d}$ is the number of processed detections.
An offset $\Delta$ is accepted, if $\mathrm{corr}_{\mathcal{P}, \mathcal{D}}(\Delta) > t$, leading to the condition
\begin{align}
N > \frac{N_\mathrm{d}}{2} (1 + t) \eqqcolon T \, .
\end{align}
The counter $N$ follows a binomial distribution and can be approximated by a normal distribution, i.e.,
\begin{align}
N \sim B(N_\mathrm{d}, p) \approx \mathcal{N}(\mu, \sigma^2) \, ,
\end{align}
with $\mu = N_\mathrm{d} \, p$, and $\sigma^2 = N_\mathrm{d} \, p (1-p)$, where $p$ is the probability for a detection to coincide with a transmitted pulse.

\subsection{Probability for accepting a wrong offset}

For wrong offsets, the coincidence probability $p = 1/2$.
Hence, for a single tested wrong offset, the acceptance probability is given by
\begin{align}
P_\mathrm{no\,wrong,1}
= P(N < T)
\approx \Phi(x) \, ,
\end{align}
where $\Phi$ is the cumulative distribution function of the normal distribution, and the standardized variable
\begin{align}
x
= \frac{T - \mu}{\sigma}
= t \sqrt{N_\mathrm{d}} \, .
\end{align}

The probability for accepting no wrong offset during the evaluation of one resynchronization block is then lower bounded by \footnote{We use a lower bound instead of equality, because acceptance of the correct offset typically terminates the search early, i.e., before covering the full search space $\Delta \in \{-\Delta_\mathrm{max}, \dots, \Delta_\mathrm{max} \}$, which effectively increases $P_\mathrm{no\,wrong}$.}
\begin{align}
P_\mathrm{no\,wrong} \ge \left( P_\mathrm{no\,wrong,1} \right)^{2 \Delta_\mathrm{max}} \, ,
\end{align}
where $\Delta_\mathrm{max}$ is the maximum tested offset.

\subsection{Probability for accepting the correct offset}

For the correct offset, the coincidence probability $p = 1 - Q$ depends on the QBER $Q$.
This leads to an acceptance probability of
\begin{align}
P_\mathrm{correct}
= P(N > T)
\approx \Phi(x) \, ,
\end{align}
with the standardized variable
\begin{align}
x
= \frac{\mu - T}{\sigma}
= \frac{\sqrt{N_\mathrm{d}}}{\sqrt{4Q(1-Q)}} (1 - t - 2 Q) \, .
\end{align}
This directly implies the condition
\begin{align}
\label{eq:threshold_condition}
t < 1 - 2 Q \, ,
\end{align}
because the success probability $P_\mathrm{correct}$ only increases with $N_\mathrm{d}$ when Eq.~\eqref{eq:threshold_condition} holds.

\subsection{Implications for the parameter choice}

The success and failure probabilities derived above allow for the choice of suitable parameters.
An acceptance threshold $t \approx 0.5$ and $N_\mathrm{d} \approx 300$ detections suffice to achieve a probability
\begin{align}
\label{eq:prob_level_1}
P_\mathrm{no\,wrong}^{(\mathrm{1\,day})} \geq 1 - 10^{-6}
\end{align}
for accepting no wrong offset during one day, assuming
an interval between successive resynchronization blocks of $I = 1 \, \mathrm{s}$,
and a probability of
\begin{align}
\label{eq:prob_level_2}
P_\mathrm{correct} \geq 1 - 10^{-2}
\end{align}
for accepting the correct offset with one resynchronization block, even for QBERs as high as $Q_\mathrm{max} = 20\,\%$.
These parameters should be suitable for most application scenarios.
In the following, the influence of the parameter choice is described in more detail.

The influence of the accepted probability levels is shown in Fig.~\ref{fig:contour_plots}(a), where the gray shaded area marks the regime with sufficiently reliable operation for most scenarios, i.e., fulfilling Eqs.~\eqref{eq:prob_level_1} and \eqref{eq:prob_level_2}.
The black dots mark the optimum solution for a particular set of requirements.
An increase of $P_\mathrm{no\,wrong}^{(\mathrm{1\,day})}$ ($P_\mathrm{correct}$) shifts the optimum solution toward higher (lower) values of the cross-correlation acceptance threshold $t$ and to a higher number of required detections $N_\mathrm{d}$, corresponding to longer resynchronization blocks.

The influence of the QBER on the parameter choice is visualized in Fig.~\ref{fig:contour_plots}(b).
Tolerance against higher QBERs is achieved by reducing $t$ and increasing $N_\mathrm{d}$.
Fulfilling Eqs.~\eqref{eq:prob_level_1} and \eqref{eq:prob_level_2} for QBERs as high as 20\,\% requires $t \approx 0.5$ and $N_\mathrm{d} \gtrsim 300$, see Fig.~\ref{fig:contour_plots}(b).
The parameter choice for the experiments (Sec.~\ref{sec:experiment}) is marked by a cross.

The influence of the maximum tested offset $\Delta_\mathrm{max}$ is shown in Fig.~\ref{fig:contour_plots}(c).
Increasing $\Delta_\mathrm{max}$ leads to a shift of the optimal solution toward increased values of $t$ and $N_\mathrm{d}$.

\begin{figure}[t]
    \includegraphics[width=0.8\linewidth]{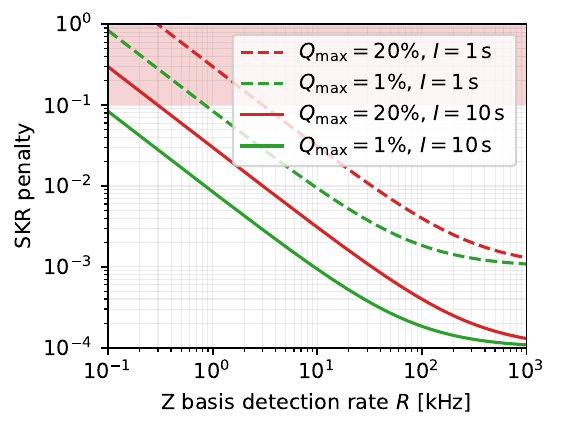}
    \caption{\label{fig:skr_impact}
    Secure-key rate (SKR) penalty during uninterrupted system operation.
    The curves are obtained by choosing the ideal values for $t$ and $N_\mathrm{d}$ for a given max. QBER $Q_\mathrm{max}$ and detection rate $R$, such that $P_\mathrm{no\,wrong}^{(\mathrm{1\,day})} = 1 - 10^{-6}$, $P_\mathrm{correct} = 1 - 10^{-2}$, and a max. tested fiber length change of $\Delta_\mathrm{max} \tau v = 100\,\mathrm{km}$, where $\tau = 800 \, \mathrm{ps}$ is the time-bin duration, and $v \approx 2.04 \times 10^8 \, \mathrm{m/s}$ is the speed of light in the fiber.
    These solutions correspond to the intersections in Fig.~\ref{fig:contour_plots}(b).
    The qubit block size $N_\mathrm{q}$ then only depends on $N_\mathrm{d}$, $R$, and the resynchronization interval $I$, directly leading to the SKR penalty.
    The red shaded area marks the regime in which the SKR is reduced by more than 10\,\%, which might be too high for many applications.
    }
\end{figure}

The impact on the SKR, due to a reduction of the qubit transmission duty cycle, is depicted in Fig.~\ref{fig:skr_impact}.
We define the SKR penalty as $N_\mathrm{r} / (N_\mathrm{q} + N_\mathrm{r})$, i.e., the temporal fraction during which the quantum channel is used for the transmission of resynchronization blocks.
For example, when transmitting a resynchronization block every $10\,\mathrm{s}$, while allowing for a QBER as high as $20\,\%$, the detection rate can be as low as $3\,\mathrm{kHz}$ while still keeping the SKR penalty below 1\,\%.
Note that this is comparable to the fraction occupied by the header (20 bytes) of TCP packets (1520 bytes), resulting in $\approx 1.3\,\%$ for the header.
Alternatively, the SKR penalty of 1\,\% can be seen as equivalent to an increase of the quantum channel attentuation of approx. $0.044\,\mathrm{dB}$, which is a negligible loss in deployed fiber networks.

\section{\label{app:security}Security considerations}

To not compromise QKD system security, the resynchronization method must not introduce any new side channels
\cite{ImplementationAttacksQKD2023}.
In the following, we identify two effects that can potentially lead to unintended information leakage, and we argue, that this can be mitigated.

To prepare the correct mean photon number for each qubit, as required by the security proof, the sender (Alice) of prepare-and-measure QKD systems has to actively control the mean photon number of the transmitted qubits.
This is typically achieved by monitoring and adjusting the optical power using a variable optical attenuator (VOA) and a PIN diode, see the \textit{PNR module} (upper purple box inside Alice) in Fig.~\ref{fig:setup}.
As feedback, the PNR uses the optical power averaged over comparatively long time intervals, e.g., 10\,ms (more than $10^6$ qubits).
The first potential side channel stems from the potentially different mean optical powers entering the PNR module for qubit blocks and resynchronization blocks.
In case the optical power entering the PNR module were lower for resynchronization blocks than for qubit blocks, the PNR module would not apply enough attenuation during the beginning of each qubit block, causing too high mean photon numbers.
To mitigate this, the average optical power entering the PNR module for resynchronization blocks must be higher than or equal to that of the qubit blocks.

A second potential side channel arises from the AC-coupling of the electrical amplifiers (see lower purple box inside Alice in Fig.~\ref{fig:setup}) used to boost signals from the FPGA to the levels required by intensity and phase modulator.
These amplifiers typically act as high-pass filters, e.g., with a 3-dB cutoff near $100\,\mathrm{kHz}$.
Consequently, the amplifier's baseline and peak output reflect a weighted average of its past input.
To avoid biasing the first qubits in each qubit block, the average electrical input level at the amplifiers must be the same for both qubit and resynchronization blocks.
Alternatively, detections for the first few qubits of each qubit block could be removed during post-processing.

\section{\label{app:nomenclature}Nomenclature}

\begin{table}
    \caption{\label{tab:nomenclature}
    Nomenclature used throughout the text.
    }
    \begin{ruledtabular}
        \begin{tabular}{p{0.20\linewidth}|p{0.77\linewidth}}
            \makecell{Symbol} & \makecell{Meaning} \\
            \colrule
            $\tau$ & Timebin duration \\
            $\mathcal{P}$ & Transmitted pattern in timebin prec., where 1 (0) represents a timebin with (without) a pulse \\
            $\mathcal{D}^{(s)}$ & Detection timestamps with sub-time-bin prec. \\
            $\mathcal{D}$ & Detection timestamps with time-bin prec. \\
            $N_\mathrm{q}$ & Length of a qubit block in timebins \\
            $N_\mathrm{r}$ & Length of a resynchronization block in timebins \\
            $N_\mathrm{d,i}$ & Number of detections for the $i$'th resync block \\
            $N_\mathrm{d,max}$ & Max. detections to process per resync block \\
            $N_\mathrm{d}$ & Processed number of detections for one resync block (after removing margins and truncating) \\
            $N$ & Counter of coinciding detections \\
            $t$ & Cross-correlation acceptance threshold \\
            $T = N_\mathrm{d} t$ & Absolute cross-correlation threshold \\
            $\mathrm{corr}$ & Cross-correlation \\
            $C = N_\mathrm{d} \mathrm{corr}$ & Absolute cross-correlation \\
            $\delta_\mathrm{align}^{(\mathrm{s})}$ & Alignment offset with sub-time-bin prec. \\
            $\Delta$ & Tested offset in timebin precision \\
            $\Delta_\mathrm{max}$ & Maximum offset to test \\
            $\delta_\mathrm{tot}^{(\mathrm{s})}$ & Total identified offset in sub-time-bin precision \\
            $B$ & Binomial distribution \\
            $\mathcal{N}$ & Standard normal distribution \\
            $p$ & Pulse coincidence probability \\
            $\mu$ & Cross-correlation expectation value \\
            $\sigma$ & Cross-correlation standard deviation \\
            $x$ & Standardized variable \\
            $I$ & Resynchronization interval \\
            $Q$ & QBER \\
            $P_\mathrm{no\,wrong}$ & Probability for accepting no wrong offset during the evaluation of one resync. block \\
            $P_\mathrm{no\,wrong}^{(\mathrm{1\,day})}$ & Probability for accepting no wrong offset during one day, given $I=1\,\mathrm{s}$ \\
            $P_\mathrm{correct}$ & Probability for accepting the correct offset during the evaluation of one resync. block \\
        \end{tabular}
    \end{ruledtabular}
\end{table}

\bibliography{literature}

\begin{thebibliography}{10}
\expandafter\ifx\csname url\endcsname\relax
  \def\url#1{\texttt{#1}}\fi
\expandafter\ifx\csname urlprefix\endcsname\relax\def\urlprefix{URL }\fi
\providecommand{\bibinfo}[2]{#2}
\providecommand{\eprint}[2][]{\url{#2}}

\bibitem{bennettQuantumCryptographyPublic1984}
\bibinfo{author}{Bennett, C.~H.} \& \bibinfo{author}{Brassard, G.}
\newblock \bibinfo{title}{Quantum cryptography: {{Public}} key distribution and coin tossing}.
\newblock \emph{\bibinfo{journal}{Proceedings of the IEEE International Conference on Computers, Systems and Signal Processing}} \bibinfo{pages}{175--179} (\bibinfo{year}{1984}).

\bibitem{gisinQuantumCryptography2002}
\bibinfo{author}{Gisin, N.}, \bibinfo{author}{Ribordy, G.}, \bibinfo{author}{Tittel, W.} \& \bibinfo{author}{Zbinden, H.}
\newblock \bibinfo{title}{Quantum cryptography}.
\newblock \emph{\bibinfo{journal}{Reviews of Modern Physics}} \textbf{\bibinfo{volume}{74}}, \bibinfo{pages}{145--195} (\bibinfo{year}{2002}).
\newblock \urlprefix\url{https://link.aps.org/doi/10.1103/RevModPhys.74.145}.

\bibitem{pirandolaAdvancesQuantumCryptography2020}
\bibinfo{author}{Pirandola, S.} \emph{et~al.}
\newblock \bibinfo{title}{Advances in {{Quantum Cryptography}}}.
\newblock \emph{\bibinfo{journal}{Advances in Optics and Photonics}} \textbf{\bibinfo{volume}{12}}, \bibinfo{pages}{1012} (\bibinfo{year}{2020}).

\bibitem{boaronSecureQuantumKey2018}
\bibinfo{author}{Boaron, A.} \emph{et~al.}
\newblock \bibinfo{title}{Secure {{Quantum Key Distribution}} over 421 km of {{Optical Fiber}}}.
\newblock \emph{\bibinfo{journal}{Physical Review Letters}} \textbf{\bibinfo{volume}{121}}, \bibinfo{pages}{190502} (\bibinfo{year}{2018}).

\bibitem{gottesmanSecurityQuantumKey2004}
\bibinfo{author}{Gottesman, D.}, \bibinfo{author}{{Hoi-Kwong Lo}}, \bibinfo{author}{L{\"u}tkenhaus, N.} \& \bibinfo{author}{Preskill, J.}
\newblock \bibinfo{title}{Security of quantum key distribution with imperfect devices}.
\newblock In \emph{\bibinfo{booktitle}{International {{Symposium}} on {{Information Theory}}, 2004. {{ISIT}} 2004. {{Proceedings}}.}}, \bibinfo{pages}{135--135} (\bibinfo{publisher}{IEEE}, \bibinfo{address}{Chicago, Illinois, USA}, \bibinfo{year}{2004}).
\newblock \urlprefix\url{http://ieeexplore.ieee.org/document/1365172/}.

\bibitem{agrawalFiberopticCommunicationSystems2010}
\bibinfo{author}{Agrawal, G.~P.}
\newblock \emph{\bibinfo{title}{Fiber-Optic Communication Systems}}.
\newblock Wiley Series in Microwave and Optical Engineering (\bibinfo{publisher}{Wiley}, \bibinfo{address}{Hoboken, NJ}, \bibinfo{year}{2010}), \bibinfo{edition}{4.} edn.

\bibitem{krauseFlexibleRealTimeQuantum2024}
\bibinfo{author}{Krause, J.}, \bibinfo{author}{Walenta, N.}, \bibinfo{author}{Hilt, J.} \& \bibinfo{author}{Freund, R.}
\newblock \bibinfo{title}{A {{Flexible Real-Time Quantum Key Distribution System}} for {{Fiber}} and {{Free-Space Links}}}.
\newblock In \emph{\bibinfo{booktitle}{{{ECOC}} 2024; 50th {{European Conference}} on {{Optical Communication}}}}, \bibinfo{pages}{23--26} (\bibinfo{year}{2024}).
\newblock \urlprefix\url{https://ieeexplore.ieee.org/document/10926438}.

\bibitem{krauseClockoffsetRecoverySublinear2025}
\bibinfo{author}{Krause, J.}, \bibinfo{author}{Walenta, N.}, \bibinfo{author}{Hilt, J.} \& \bibinfo{author}{Freund, R.}
\newblock \bibinfo{title}{Clock-offset recovery with sublinear complexity enables synchronization on low-level hardware for quantum key distribution}.
\newblock \emph{\bibinfo{journal}{Physical Review Applied}} \textbf{\bibinfo{volume}{23}}, \bibinfo{pages}{044015} (\bibinfo{year}{2025}).

\bibitem{martinMadQCIHeterogeneousScalable2024}
\bibinfo{author}{Martin, V.} \emph{et~al.}
\newblock \bibinfo{title}{{{MadQCI}}: A heterogeneous and scalable {{SDN-QKD}} network deployed in production facilities}.
\newblock \emph{\bibinfo{journal}{npj Quantum Information}} \textbf{\bibinfo{volume}{10}}, \bibinfo{pages}{80} (\bibinfo{year}{2024}).

\bibitem{aliaDynamicDVQKDNetworking2022}
\bibinfo{author}{Alia, O.} \emph{et~al.}
\newblock \bibinfo{title}{Dynamic {{DV-QKD Networking}} in {{Trusted-Node-Free Software-Defined Optical Networks}}}.
\newblock \emph{\bibinfo{journal}{Journal of Lightwave Technology}} \textbf{\bibinfo{volume}{40}}, \bibinfo{pages}{5816--5824} (\bibinfo{year}{2022}).

\bibitem{simSoftwareDefinedNetworkingOrchestration2023}
\bibinfo{author}{Sim, D.-H.}, \bibinfo{author}{Shin, J.} \& \bibinfo{author}{Kim, M.~H.}
\newblock \bibinfo{title}{Software-{{Defined Networking Orchestration}} for {{Interoperable Key Management}} of {{Quantum Key Distribution Networks}}}.
\newblock \emph{\bibinfo{journal}{Entropy}} \textbf{\bibinfo{volume}{25}}, \bibinfo{pages}{943} (\bibinfo{year}{2023}).

\bibitem{bhatiaDynamicSecurityAwareResource2025}
\bibinfo{author}{Bhatia, V.}, \bibinfo{author}{Kasegenya, A.} \& \bibinfo{author}{Chen, B.}
\newblock \bibinfo{title}{Dynamic {{Security-Aware Resource Allocation}} in {{Quantum Key Distribution-Enabled Optical Networks}}}.
\newblock \emph{\bibinfo{journal}{Photonics}} \textbf{\bibinfo{volume}{12}}, \bibinfo{pages}{645} (\bibinfo{year}{2025}).
\newblock \urlprefix\url{https://www.mdpi.com/2304-6732/12/7/645}.

\bibitem{selentis-boulntadakisRelayedVsSwitched2025}
\bibinfo{author}{{Selentis-Boulntadakis}, A.}, \bibinfo{author}{Christodoulopoulos, K.} \& \bibinfo{author}{Kanellos, G.~T.}
\newblock \bibinfo{title}{Relayed vs. {{Switched QKD}}: {{A Comparison}} in {{Non-Uniform Ring Networks}}}.
\newblock In \emph{\bibinfo{booktitle}{2025 {{International Conference}} on {{Optical Network Design}} and {{Modeling}} ({{ONDM}})}}, \bibinfo{pages}{1--6} (\bibinfo{year}{2025}).
\newblock \urlprefix\url{https://ieeexplore.ieee.org/document/11029349/}.

\bibitem{ruscaFinitekeyAnalysis1decoy2018}
\bibinfo{author}{Rusca, D.}, \bibinfo{author}{Boaron, A.}, \bibinfo{author}{Gr{\"u}nenfelder, F.}, \bibinfo{author}{Martin, A.} \& \bibinfo{author}{Zbinden, H.}
\newblock \bibinfo{title}{Finite-key analysis for the 1-decoy state {{QKD}} protocol}.
\newblock \emph{\bibinfo{journal}{Applied Physics Letters}} \textbf{\bibinfo{volume}{112}}, \bibinfo{pages}{171104} (\bibinfo{year}{2018}).

\bibitem{greggSystemsPerformanceEnterprise2021}
\bibinfo{author}{Gregg, B.}
\newblock \emph{\bibinfo{title}{Systems Performance: {{Enterprise}} and the {{Cloud}}}}.
\newblock Addison-{{Wesley}} Professional Computing Series (\bibinfo{publisher}{Addison-Wesley}, \bibinfo{year}{2021}), \bibinfo{edition}{2.} edn.

\bibitem{cochranQubitBasedClockSynchronization2021}
\bibinfo{author}{Cochran, R.~D.} \& \bibinfo{author}{Gauthier, D.~J.}
\newblock \bibinfo{title}{Qubit-{{Based Clock Synchronization}} for {{QKD Systems Using}} a {{Bayesian Approach}}}.
\newblock \emph{\bibinfo{journal}{Entropy}} \textbf{\bibinfo{volume}{23}}, \bibinfo{pages}{988} (\bibinfo{year}{2021}).

\bibitem{Note1}
\bibinfo{note}{We use a lower bound instead of equality, because acceptance of the correct offset typically terminates the search early, i.e., before covering the full search space $\Delta \in \{-\Delta _\protect \mathrm {max}, \protect \dots , \Delta _\protect \mathrm {max} \}$, which effectively increases $P_\protect \mathrm {no\protect \,wrong}$.}

\bibitem{ImplementationAttacksQKD2023}
\bibinfo{title}{Implementation {{Attacks}} against {{QKD Systems}}}.
\newblock \bibinfo{type}{Tech. Rep.}, \bibinfo{institution}{Bundesamt f{\"u}r Sicherheit in der Informationstechnik (BSI)} (\bibinfo{year}{2023}).
\newblock \urlprefix\url{https://www.bsi.bund.de/EN/Service-Navi/Publikationen/Studien/QKD-Systems/Implementation_Attacks_QKD_Systems_node.html}.

\end{thebibliography}

\end{document}